\documentclass{iopjournal}

\usepackage{amssymb}
\usepackage{xcolor}
\usepackage{hyperref}
\usepackage{upgreek}
\usepackage{graphicx}
\usepackage{subcaption}
\usepackage{amsmath}
\usepackage{textcomp}
\usepackage[percent]{overpic}
\usepackage{ragged2e}
\usepackage{microtype}
\usepackage{lmodern}  

\makeatletter
\let\@copyrightline\relax
\makeatother

\makeatletter
\let\@vtext\@empty          
\let\@oddhead\@empty        
\let\@evenhead\@empty       
\makeatother

\begin{document}
\sloppy
\justifying

\articletype{xxxx} 

\title{Toward Quantitative Electric-Field Measurements of Inception Clouds in Nanosecond Discharges Using E-FISH Assisted by Machine Learning}

\author{Mhedine Alicherif$^{1,2,*}$\orcid{0009-0003-9286-8703}, Edwin Sugeng$^3$\orcid{0000-0002-7175-2467}, Yang Zhijan$^3$\orcid{0000-0002-8287-4267}, Deanna A. Lacoste$^{1,2}$\orcid{0000-0002-4160-4762}  and Tat Loon Chng$^{3}$\orcid{0000-0001-7621-1750}}

\affil{$^1$Mechanical Engineering Program, Physical Science and Engineering Division, King Abdullah University of Science and Technology (KAUST), Thuwal, Saudi Arabia}

\affil{$^2$Clean Energy Research Platform, King Abdullah University of Science and Technology, Thuwal, Saudi Arabia}

\affil{$^3$Department of Mechanical Engineering, College of Design and Engineering, National University of Singapore, 117575, Singapore}

\affil{$^*$Author to whom any correspondence should be addressed.}

\email{mhedine.alicherif@kaust.edu.sa}


\begin{abstract}
This study investigates the spatio–temporal evolution of the electric field during the early stages of a nanosecond positive corona discharge in atmospheric-pressure air by combining time-resolved E-FISH measurements, machine-learning-assisted field inversion (based on a recently developed operator-learning model), and iCCD optical emission imaging. The objective is to quantitatively characterize the electric field in the vicinity of the high-voltage electrode during inception and the transition toward streamer formation. By averaging over a large number of discharge events and operating in a regime where the discharge remains statistically axisymmetric, the proposed approach enables reconstruction of the electric-field profiles with nanosecond resolution. The results show a rapid increase of the field during the first nanoseconds, followed by the formation of a shell-like structure exhibiting the highest reduced fields prior to destabilization. The reconstructed reduced electric-field magnitude reaches peak values in the range of approximately 230–270 Td, with an estimated uncertainty of about 20–30\% associated with calibration and profile-shape effects. These values correspond to the regime where electron-impact excitation and photoionization processes become highly efficient, consistent with the observed transition from a stable inception cloud to streamer destabilization. After the onset of streamer branching, increasing asymmetry limits the applicability of the inversion, and the reconstructed fields represent averaged contributions rather than the local field at individual streamer heads. The methodology thus identifies the conditions under which quantitative E-field mapping is reliable and establishes a framework for extending electric-field diagnostics to the inception phase of nanosecond atmospheric discharges.
\end{abstract}

\section{Introduction}

Nanosecond discharges are characterized by a rapid voltage rise and short duration, producing highly energetic electrons while minimizing thermal load~\cite{wang2020nanosecond}. These conditions promote non-equilibrium chemistry, enhancing reactivity at relatively low energy cost. As a result, they are particularly relevant for applications such as plasma-assisted combustion~\cite{lacoste2023flames, starikovskaia2014plasma, popov2016kinetics}, water and liquid treatment~\cite{kolb2008streamers}, industrial surface modification~\cite{bardos2010cold}, and ozone generation~\cite{van2008evaluation}.

Within this family of discharges, corona discharges occur when the electric field is sufficiently high to ionize the surrounding fluid near a conductor, without immediate arc formation. These discharges are of particular interest not only for their industrial implications, ranging from detrimental effects such as energy losses in high-voltage power lines and radio interference, to beneficial uses like ozone production and surface treatment, but also because they represent a critical stage in the potential transition toward arc formation.

The temporal evolution of a positive corona discharge proceeds through several distinct phases: an initial homogeneous inception cloud, followed by the formation of a shell structure, its destabilization, and ultimately the emergence of filamentary streamer heads~\cite{nijdam2020physics}. Understanding the mechanisms driving this destabilization is essential for both predictive modeling and application-specific control. Central to this evolution is the electric field, which governs electron avalanches, photon production for photoionization, and streamer propagation. Accurate characterization of the electric field is therefore critical to understanding the physics underlying this transition.

Measuring the electric field in such dynamic and spatially complex environments is inherently challenging. Intrusive probes such as electrostatic or electro-optic sensors are unsuitable due to the small spatial scales involved and the risk of perturbing the discharge. Optical techniques like Stark broadening~\cite{takiyama1986measurement, kuraica1997electric} and intensity-ratio-based optical emission spectroscopy~\cite{obrusnik2018electric, bilek2018electric, bonaventura2011electric, paris2005intensity} offer non-intrusive alternatives, but suffer from limitations including limited spatial resolution, line-of-sight integration, and dependence on collisional-radiative models. Electric Field Induced Second Harmonic Generation (E-FISH) has emerged as a powerful laser-based diagnostic capable of probing electric fields at atmospheric pressure with high temporal resolution. While it has been successfully applied in various discharge conditions~\cite{simeni2018electricz, chng2019electric, chng2019electricz, orr2020measurements, adamovich2020nanosecond, simeni2018electric, goldberg2018electric, lepikhin2020electric}, quantitative measurements in nanosecond positive corona discharges remain limited.

A key limitation of E-FISH is that the signal represents an interaction of the focused laser beam with the surrounding electric field along its entire optical path, including regions that may lie well beyond the focal volume~\cite{chng2020electric}. As a result, an E-FISH measurement at a single point is not only dependent on the local applied electric field at the beam focus, but also the shape of this entire profile along the axis of laser propagation. More details of this are discussed in section~\ref{Data}. One possible solution is to perform spatial scans of the E-FISH signal along the laser axis and then reconstruct the corresponding field profile from these scans. This latter process, in our opinion, poses a non-trivial inversion problem that requires sophisticated reconstruction techniques to retrieve the underlying electric field shape from the integrated signal profile.

The advent of machine learning (ML) algorithms has offered a novel and yet promising approach to the E-FISH inversion task described above. The fundamental idea involves training a model on a sufficiently large and representative set of E-FISH and electric field profile pairs. This process allows the model to learn the complex, nonlinear relationship between an electric field distribution and E-FISH signal profile. Such an ML approach offers several key advantages. First, once a model has been trained and tested, it is often highly computationally efficient, enabling near real-time predictions of electric field profiles. Furthermore, the performance of such a model can be readily adapted or refined by expanding upon its training dataset, for instance, when new data or findings become available. Finally, since the underlying training data can be efficiently synthesized from theory (via equation (2)), there is no fundamental restriction to generating these large amounts of information, thus overcoming the scarcity limitation that often restricts the performance of such ML models.

In line with the above, an ML model based on a convolutional neural network (CNN) has demonstrated excellent predictive accuracy and noise-resilience when applied to synthetic E-FISH profiles, as well as experimental electrostatic fields obtained from spatial, 1-D E-FISH signal scans ~\cite{yang2025deep}. However, a perceived limitation of this model is that it is trained on a specific electric field profile shape, which may affect its ability to generalize to (i.e., accurately predict) more complex profiles. More recently, a new model built on a more sophisticated neural operator architecture (DDON) has exhibited superior performance (compared to the CNN) both in terms of accuracy and generalizability, as well as when applied to experimental data acquired in a discharge~\cite{Yang2025Interpretable}. A unique feature of neural operators (on which DDON is developed) is that they are specifically designed for learning more complex function-to-function mappings, thus addressing the foregoing issue of generalizability faced by conventional CNNs. It is also less sensitive to the exact location(s) of the acquired data, enabling electric field reconstruction even with seemingly ‘incomplete’ input profiles – an issue which often accompanies poor signal sensitivity in real-life experiments.

With the above in mind, the present study proposes and validates a diagnostic approach suitable for quantitative measurements of the electric field in a corona discharge, specifically targeting the transition from the homogeneous inception cloud to the formation of streamer heads. This approach leverages our existing DDON model for E-FISH profile inversion to measure the electric field magnitude, and combines this with iCCD imaging to interpret the reconstructed field profiles and visualize the results. The inclusion of iCCD imaging as a complementary, evidence-based tool permits an independent evaluation of the E-FISH data, particularly under stochastic discharge conditions.

Section~\ref{Expe} presents the experimental setup and diagnostics. The ML reconstruction methodology is detailed in Section~\ref{Data}. Results from iCCD emission imaging, E-FISH measurements, and electric field reconstruction are reported in Section~\ref{Res}. Finally, Section~\ref{Disc} offers a critical analysis of the results, with particular attention to the limitations inherent to both the E-FISH technique and the corona discharge configuration.

\section{Experimental Setup}
\label{Expe}
\begin{figure}[htbp]
    \centering

        \centering
        \includegraphics[width=\textwidth]{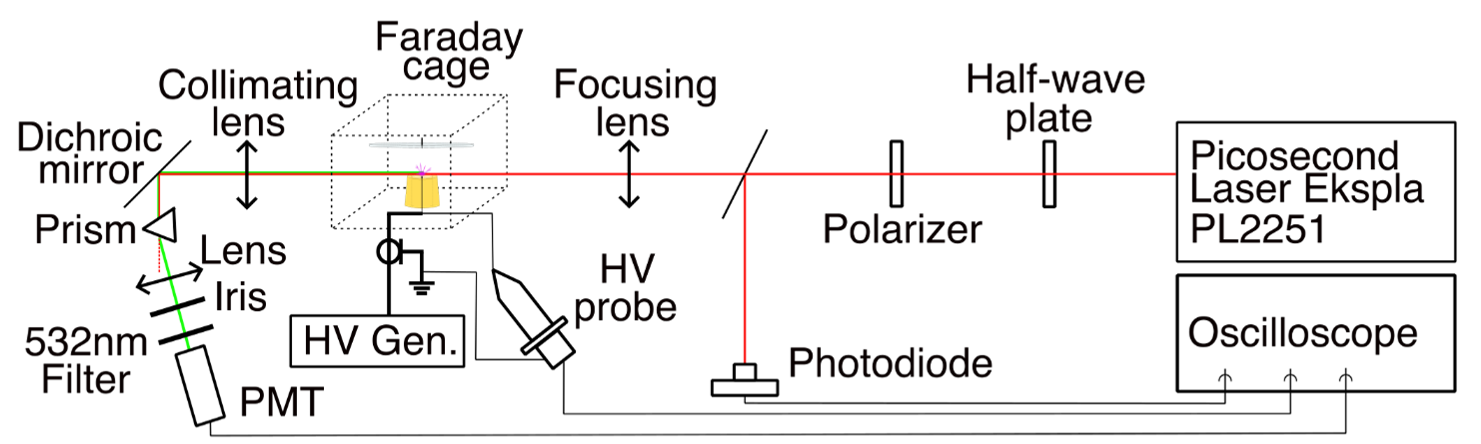}
    \caption{Schematic of E-FISH Experimental setup. HV=High-voltage; Gen.=Generator.}
    \label{fig:setup}
\end{figure}

Figure~\ref{fig:setup} illustrates the experimental setup and Figure~\ref{fig:em} provides a magnified view of the electrode configuration with scanning electron microscope images of the electrode tips.

\subsection{Specifics of the plasma}
\label{Specplasma}

The reactor used in this work is essentially based on a stagnation flame burner designed for plasma-assisted combustion research, which has been well-described in previous work~\cite{alkhalifa2024quantifying}. The setup is operated in the absence of any flame, i.e., in ambient air.

\begin{figure}[htbp]
    \centering
        \centering
        \includegraphics[width=\textwidth]{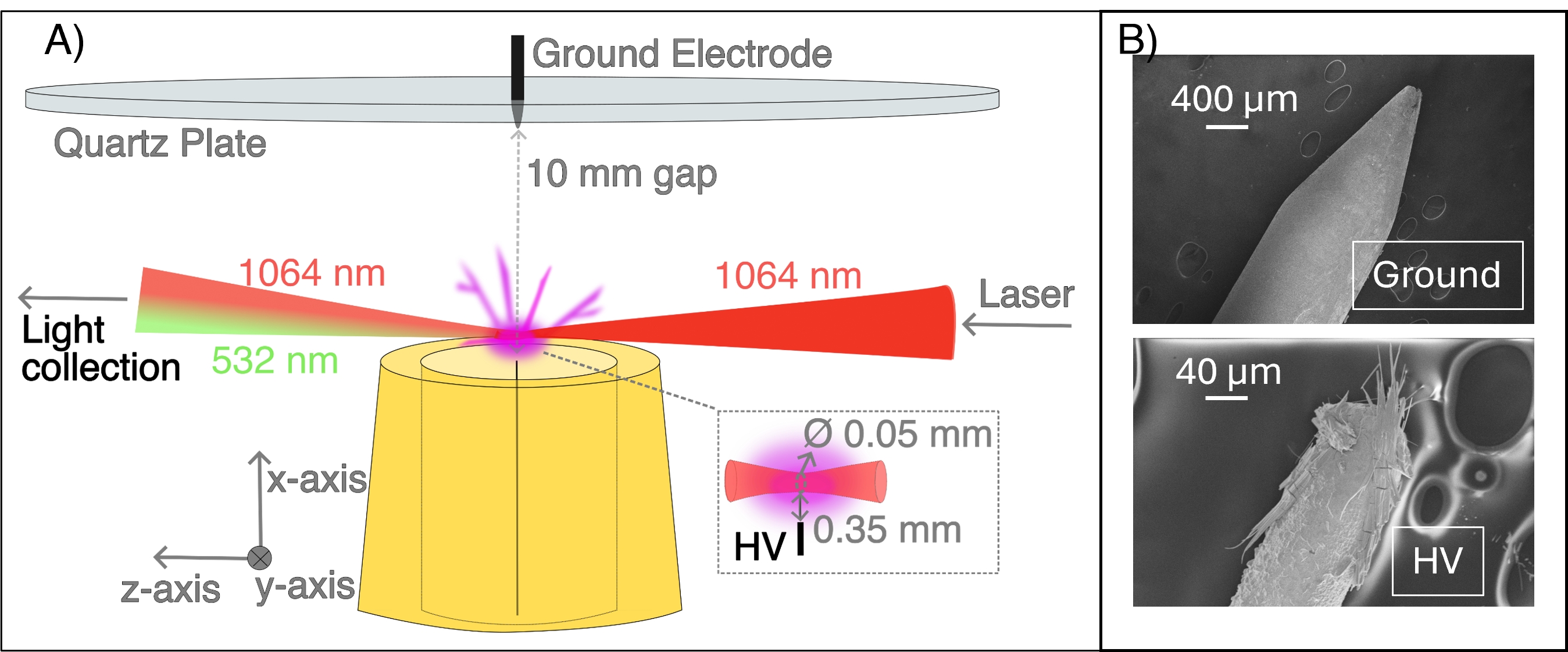}
    \caption {A) Magnified schematic of the reactor and electrode configuration together with B) Scanning electron microscope images of the electrode tips.}
    \label{fig:em}
\end{figure}

Briefly, the electrode assembly consists of two tungsten electrodes separated by a 10~mm vertical gap. The grounded electrode (Fig.~\ref{fig:em}(B)) is needle-shaped with a tip diameter of 160~\textmu m, while the high-voltage electrode (Fig.~\ref{fig:em}(B)) is a cylindrical rod with a tip diameter of 111~\textmu m. Both electrodes are connected to a high-voltage pulse generator (FID GmbH Model FPG~10-30NM10), delivering 10~ns (FWHM) pulses at a repetition rate of 10~kHz, with amplitudes ranging from 3.5 to 8.5~kV. The voltage between the electrodes is measured using a high-voltage probe (Tektronix P6015A, 75~MHz).

To provide a clearer view of their geometry and dimensions, scanning electron microscope (SEM) images of both electrodes are acquired and shown in Fig.~\ref{fig:em}(B). More importantly, this enables a more accurate representation when performing numerical simulations of the static electric field (see Section~\ref{ml}).

\subsection{E-FISH diagnostic}
\label{EFD}

This section provides details of the picosecond (ps) Electric-Field Induced Second Harmonic Generation (E-FISH) setup used to collect the experimental signal profiles discussed in Section~\ref{Res}. It is based on a standard configuration (Fig.~\ref{fig:setup}), similar to that described in Refs.~\cite{chng2019electric, yang2025deep}, with all experiments conducted at atmospheric pressure and ambient temperature (nominally 1~bar and 25\textdegree{}C).

A vertically polarized 1064~nm fundamental output beam from a 10~Hz, \textbf 4.2~mJ, 30~ps laser (EKSPLA PL2251) is directed at a distance of approximately 0.34~mm above the high-voltage (HV) electrode (along the \textit{x}-axis), as indicated in Fig.~\ref{fig:em}(A). The intensity of the 1064~nm beam is monitored via a fast photodiode with a rise time of 1~ns (Thorlabs DET10A2). The beam is focused using a plano-convex lens with a 25~cm focal length.

The Rayleigh range, determined using knife-edge measurements, is $z_R = 1.35$~mm, which agrees well with the Gaussian beam approximation based on an 8~mm collimated beam diameter. Combined with a wavevector mismatch $\Delta k = 0.5$~cm$^{-1}$, this yields a phase mismatch parameter $u = -\Delta k \cdot z_R \approx -0.068$.

A longpass filter (Thorlabs FGL850S), placed after the focusing lens, blocks stray second harmonic generation (SHG) occurring prior to the signal region. After traversing the field region, the beam is collimated and magnified using a second plano-convex lens with a 50~cm focal length. The E-FISH signal at 532~nm is separated from the fundamental using a dichroic mirror and a dispersive prism. This signal is then focused with a 10~cm focal length lens into a gated photomultiplier tube (PMT) module (Hamamatsu H11526-NF, 1~ns rise time).

A gating window of approximately 50~ns is applied to the PMT to limit its active period during signal acquisition, thereby reducing background plasma emission and enhancing the signal-to-noise ratio. An iris and a 532~nm bandpass filter (Edmund Optics, OD4, FWHM 10~nm) are placed at the PMT input for further stray light rejection. Both photodiode and PMT signals are recorded using a digital oscilloscope (Teledyne LeCroy T3DSO31004) with a 5~GHz sampling rate and 1~GHz bandwidth.

\begin{figure}[htbp]
    \centering
        \centering
        \includegraphics[width=0.75\textwidth]{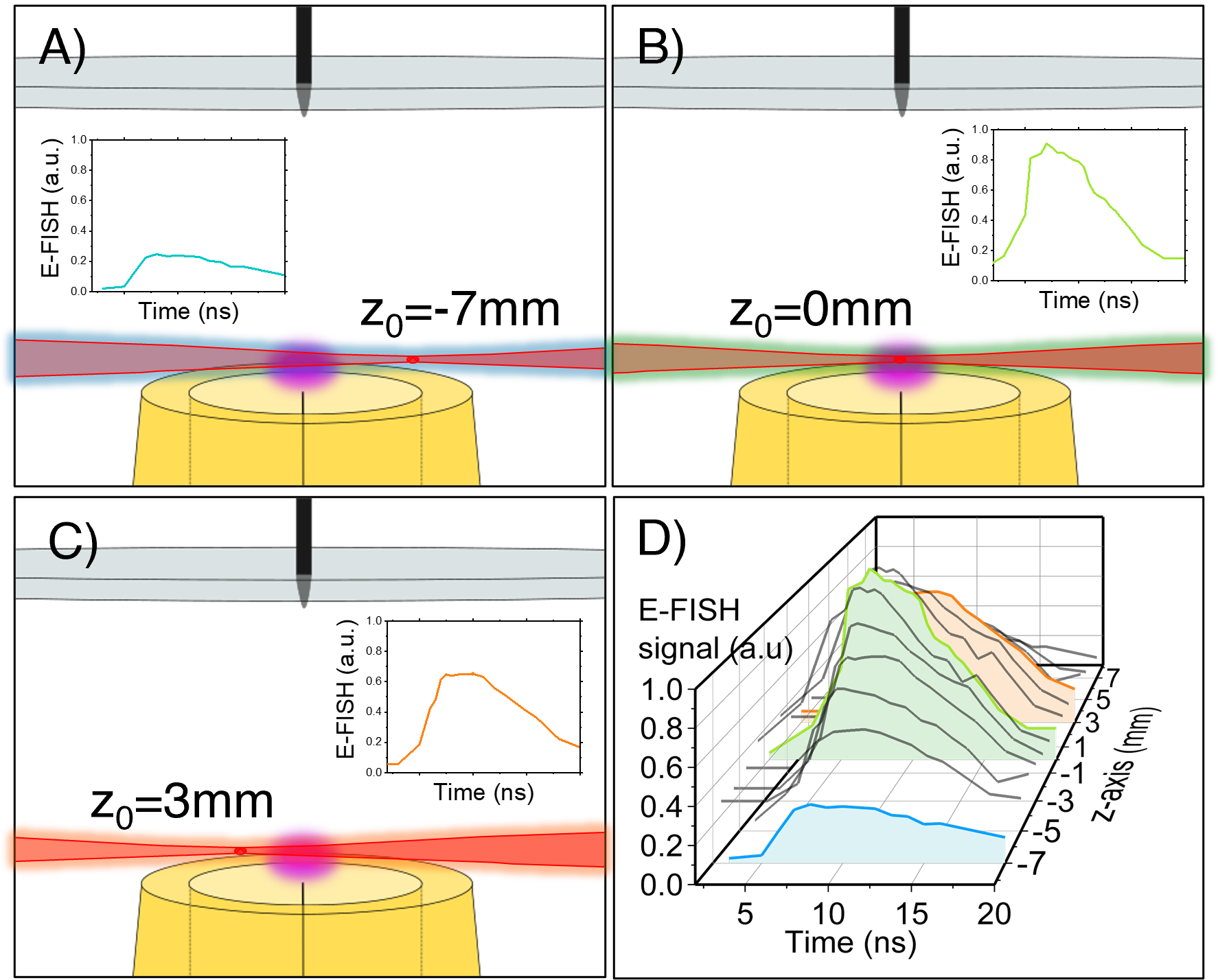}
    \caption {Schematic illustrating the E-FISH spatial scan methodology: A), B), and C), different z positions (location of beam focus is represented by a bright red spot) and D) corresponding colored E-FISH signal in the complete scan.}
    \label{fig:measure}
\end{figure}

The procedure used to measure the E-FISH signal for field reconstruction is illustrated in Fig.~\ref{fig:measure}. To obtain an E-FISH profile, a spatial scan is performed by translating the electrode assembly along the \textit{z}-axis using a 3-axis translation stage in small increments of $\Delta z = 0.5$–1~mm ($\Delta z_o' \approx 1$) over a range of  $z=\pm$20~mm. Under post-breakdown conditions, the electric field profile is additionally time-dependent, thus necessitating a spatio-temporal scan. For these cases, the E-FISH signal is recorded over a 20~ns time window at each spatial location, which fully covers the discharge duration (15~ns), and at each time instant, averaged over 2048 laser shots to improve the signal-to-noise ratio. This process is detailed in Fig.~\ref{fig:measure}(a–c), where the position of the laser beam relative to the electrodes is denoted by its coordinate $z$ (with $z = 0$~mm corresponding to the axis of symmetry). The entire spatio-temporal scan is shown in Fig.~\ref{fig:measure}(D), where colored lines correspond to the specific examples highlighted in Fig.~\ref{fig:measure}(A–C), and the remaining profiles are plotted in dark gray.
Each profile is normalized by the square of the laser intensity after background subtraction. The resulting profiles are then peak-normalized and used as inputs for the DDON model to infer $E'_{\text{ext}}(z')$.

Two pulse delay generators (Berkeley Nucleonics, BNC 577-4C and BNC 575-8C) are employed for synchronization between the laser pulse, high-voltage pulses, and the PMT gating signal. The overall timing jitter is maintained below 3~ns. Due to the laser’s repetition rate (limited to 10~Hz) being significantly lower than the discharge frequency (10~kHz), each E-FISH data point samples approximately once every 1000 HV pulses.

\subsection{Optical emission imaging}

To provide further insight into the electric field data, time-resolved optical emission images of the discharge are captured. These experiments are performed using a PiMAX4 UV camera (Princeton Instruments) in combination with a long-distance microscope (Cerco UV lens 200~mm f/2.8). The camera is operated in two modes at maximum gain - with a gate width of 20~ns to capture the whole discharge, and a gate width of 1~ns to capture the discharge dynamics.

\section{Data analysis}
\label{Data}

\subsection{Electric field profile reconstruction}
\label{calibration}

To reiterate an earlier point, one of the key objectives of this work is to apply a machine-learning (ML) algorithm (DDON) developed previously \cite{Yang2025Interpretable} for the specific purpose of electric field profile reconstruction and quantification.

The need for such a reconstruction algorithm stems from an intrinsic limitation of the E-FISH diagnostic (when focused beams are utilized), an issue which has been extensively discussed in refs.~\cite{yang2025deep, chng2020electric, chng2022effect}. It is worth mentioning that various other approaches have been proposed to address this problem \cite{nakamura2021electric, nakamura2022measurement, limburgnumerical, guo2025measurement, chen2024measurement}. Essentially, due to an additional Gouy phase shift inherent in a focused laser beam, the E-FISH signal becomes dependent upon the entire (shape of the) electric field profile that overlaps with the resulting beam path. This counterintuitive result generally runs contrary to other nonlinear laser-based methods, where the expectation is that the signal can often be localized to the confocal beam parameter (i.e., the region of highest intensity).

Following \cite{yang2025deep}, this phenomenon is mathematically captured by the equation for the E-FISH signal power:
\begin{equation}
P^{(2\omega)} \propto \left[ \alpha^{(3)} \cdot N \cdot P_o^{(\omega)} \right]^2 \cdot \frac{1}{z_R} \cdot \left| \int_{-\infty}^{\infty} \frac{E_{\text{ext}}(z) \cdot e^{i \Delta k z}}{1 + i z/z_R} \, dz \right|^2,
\end{equation}
where $\alpha^{(3)}$ is the third-order nonlinear hyperpolarizability (a fourth-rank tensor), $N$ is the gas number density, $P_o^{(\omega)}$ is the probe beam power, $z_R$ is the laser beam Rayleigh range, $E_{\text{ext}}(z)$ is the externally applied electric field distribution along the $z$-axis, $\Delta k$ is the wave-vector mismatch, and $z$ is the beam propagation axis, such that both the beam focus and the origin of $E_{\text{ext}}(z)$ are located at $z=0$. (For completeness, the Gouy phase shift is defined as $\tan^{-1}(z/z_R)$.)

It is evident from Equation~(1) that unless the underlying electric field profile, $E_{\text{ext}}(z)$, is known, the E-FISH signal cannot be accurately defined. To overcome this ill-posed relationship, the approach evaluated in this work involves acquiring an E-FISH signal profile along the axis of laser propagation (i.e., $z$-axis), facilitated by sequentially displacing (or spatially scanning) the plasma relative to the beam focus as shown in Fig.~\ref{fig:measure}. This signal profile is then used to reconstruct $E_{\text{ext}}(z)$ based on Equation~(2) below:
\begin{equation}
P^{(2\omega)}(z_o) \propto \left[ \alpha^{(3)} \cdot N \cdot P_o^{(\omega)} \right]^2 \cdot \frac{1}{z_R} \cdot \left| \int_{-\infty}^{\infty} \frac{E_{\text{ext}}(z - z_o) \cdot e^{i \Delta k z}}{1 + i z/z_R} \, dz \right|^2,
\end{equation}
where the variable $z_o$ is a measure of the displacement between the laser beam focus (located at z = 0) and the electric field profile.

Given that the process of obtaining $E_{\text{ext}}(z)$ from $P^{(2\omega)}(z_o)$ involves (in our opinion) a nontrivial mathematical inverse problem, an ML algorithm based on a neural operator architecture (i.e., DDON) is used to assist in this reconstruction. Further details of this model can be found in \cite{Yang2025Interpretable}, but a notable limitation is that it is only applicable to the reconstruction of bell-shaped or double-peak profiles with an existing axis of symmetry, and does not generalize to asymmetric as well as antisymmetric distributions, nor profiles of an entirely random nature.

This limitation invariably restricts the use of DDON to discharges with a high degree of repeatability (low stochasticity). Applying this to discharges with morphologies that are susceptible to streamer branching and filamentation phenomenon, is therefore challenging. To address this, iCCD imaging is used as a complementary, evidence-based tool for interpreting the time-averaged E-FISH data, a point we address further in Section~\ref{Disc}.

For convenience, Equation~(2) can be rewritten in its non-dimensional form as:
\begin{equation}
P^{(2\omega)}(z_o') \propto \left[ \alpha^{(3)} \cdot N \cdot P_o^{(\omega)} \right]^2 \cdot E_o^2 \cdot z_R \cdot \left| \int_{-\infty}^{\infty} \frac{E'_{\text{ext}}(z' - z_o') \cdot e^{i u z'}}{1 + i z'} \, dz' \right|^2,
\end{equation}
where
\begin{align*}
E'_{\text{ext}}(z') &= \frac{E_{\text{ext}}(z)}{E_o}, \\
\end{align*}
and $E_o$ represents the peak electric field strength while $E'_{\text{ext}}(z')$ fully captures the (normalized) shape of the underlying electric field profile. Note that the ML algorithm accepts the peak-normalized E-FISH signal profile, $P^{(2\omega)}_{\text{norm}}(z_o')$, as an input, and outputs the underlying electric field shape, $E'_{\text{ext}}(z')$.

The above notwithstanding, only the shapes of the respective electric field profiles are addressed by the DDON model, while the magnitude of these profiles (i.e., field strength) is obtained via calibration (see following Section~\ref{ml}).

\subsection{Electric field calibration}
\label{ml}

Following the previous sub-section and equation (3), it is worth pointing out that once the shape of the field profile has been obtained, only $E_o$ needs to be found before the magnitude of the entire profile is known. For this purpose, we develop a procedure that (i) incorporates standard aspects of signal calibration used in the E-FISH literature, while also (ii) accounting for the effects of profile shape via our DDON algorithm. From a mathematical point of view (based on equation (3) above), part (i) describes the process of computing $E_o$, while (ii) focuses on validating and applying $E'_{\text{ext}}(z')$ obtained via the DDON model. 

\begin{figure}[htbp]
    \centering
        \centering
        \includegraphics[width=\textwidth]{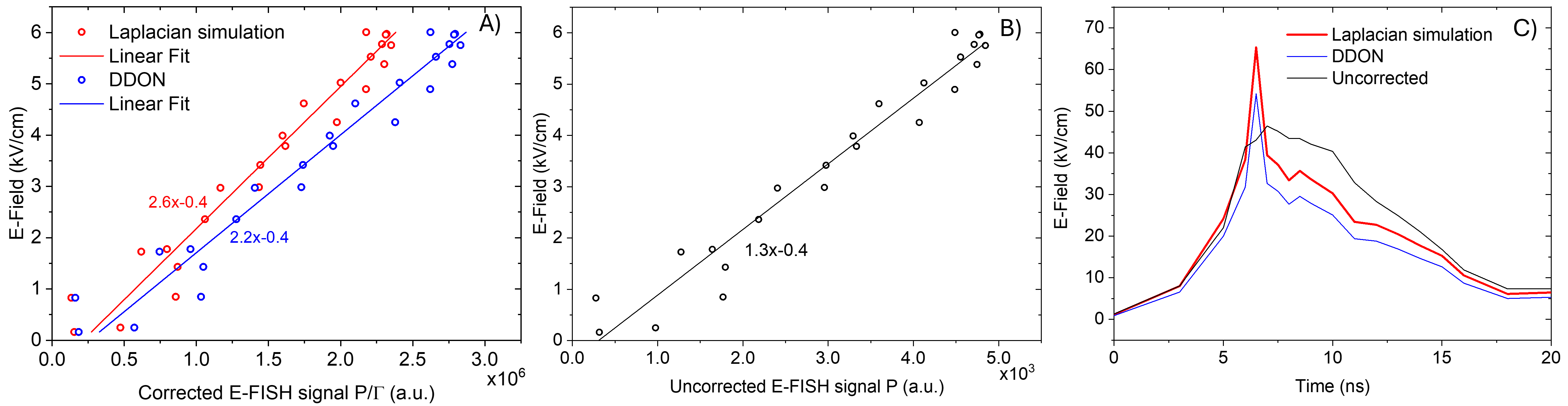}
    \caption{Plot of calibration curves generated from A) experiments versus simulation results following the procedure depicted in figure~\ref{fig:calibration} and B) regular calibration procedure typically adopted in the literature. C) Comparison of the e-field magnitudes obtained using the E-FISH signals corrected for the profile shapes using the DDON model versus the uncorrected values.}
    \label{fig:cal}
\end{figure}

As with most E-FISH experiments, signal calibration in this work is performed under electrostatic conditions (corresponding to an applied voltage of 3.5 kV), and tagged to an absolute scale using numerical simulations. The Partial Differential Equation (PDE) toolbox in MATLAB, which assumes two-dimensional (2-D) axisymmetry, is used for this purpose. These simulations of the Laplacian field model almost the exact reactor geometry including both the nozzle outlet and quartz plate surrounding the HV and ground electrode respectively. The precise shape of both electrodes is also adopted based on the SEM images presented earlier in figure~\ref{fig:em}.

These simulations serve several important purposes. The first, as with most typical E-FISH experiments, is to determine the magnitude of the peak electric field strength, or $E_o$, in equation~(3). This is essential since the electric field strength at any location within the interelectrode gap (of this pin-pin assembly) is not equal to the applied voltage (at the HV electrode) divided by this gap distance. To populate the calibration curve, changes in the E-FISH signal with field strength are obtained by simply matching the measured signal to its corresponding value within the duration of a HV nanosecond pulse~\cite{chng2019electric, chen2024measurement}. The temporal shape of the HV waveform is measured directly using the HV probe mentioned in Section~\ref{Specplasma}, and scaled to its appropriate value within the gap based on the simulation values. The two acquired sets of temporal signals (E-FISH vs HV pulse) are subsequently overlaid (in time) using a cross-correlation algorithm to yield the best fit.

The second purpose, as mentioned above, which is both unique and essential to this work, is to account for the shape of the electric field profile on the E-FISH signal. To this end, we define based on equation~(3), a ‘modification-factor’, $\Gamma$, which can be mathematically expressed following~\cite{chng2022effect} as:
\begin{equation}
\Gamma = \left| \int_{-\infty}^{\infty} \frac{E'_{\text{ext}}(z')\cdot e^{i u z'}}{1 + i z'} \mathrm{d}z' \right|^2
\end{equation}
where all terms are as previously defined. It can be seen from equation~(4) that the effect of the profile shape is fully captured by $\Gamma$, and by assuming further that optical properties such as $u$ and $z_R$ remain constant during an E-FISH experiment, one notes that $\Gamma$ is dominantly affected by $E'_{\text{ext}}(z')$.

To explicitly account for the shape of the field profile, the calibration curve proposed and utilized in this work (see figure~\ref{fig:cal}A) instead plots the peak field strength, $E_o$ versus the E-FISH signal, $P$ normalized by the modification factor, $\Gamma$. This 'corrected' E-FISH signal ($P$/$\Gamma$) serves to account for any changes in the field profile, in contrast to the regular calibration procedure typically adhered to in the literature (shown in figure~\ref{fig:cal}B for comparison).


It is highlighted here that accurate use of figure~\ref{fig:cal} requires a knowledge of the profile shape at a given measurement instant, so that $\Gamma$ may be computed. As mentioned earlier, this is especially crucial under discharge conditions, since the shape of the electric field profile is expected to evolve with time, in contrast to the electrostatic (viz. calibration) case, where the electric field profile shape is assumed to be time-invariant.

The third (and final) purpose of the simulations is that they provide a direct comparison between the electric field profile, $E'_{\text{ext}}(z')_{\text{sim}}$, against that predicted by our DDON algorithm from the experimental data, $E'_{\text{ext}}(z')_{\text{ML}}$. Under calibration (electrostatic) conditions, one expects that both these profiles should be identical. Therefore any differences between $E'_{\text{ext}}(z')_{\text{ML}}$ and $E'_{\text{ext}}(z')_{\text{sim}}$ will affect $\Gamma$, and in turn reflect the uncertainty associated with the calibration.

\begin{figure}[htbp]
    \centering

        \centering
        \includegraphics[width=\textwidth]{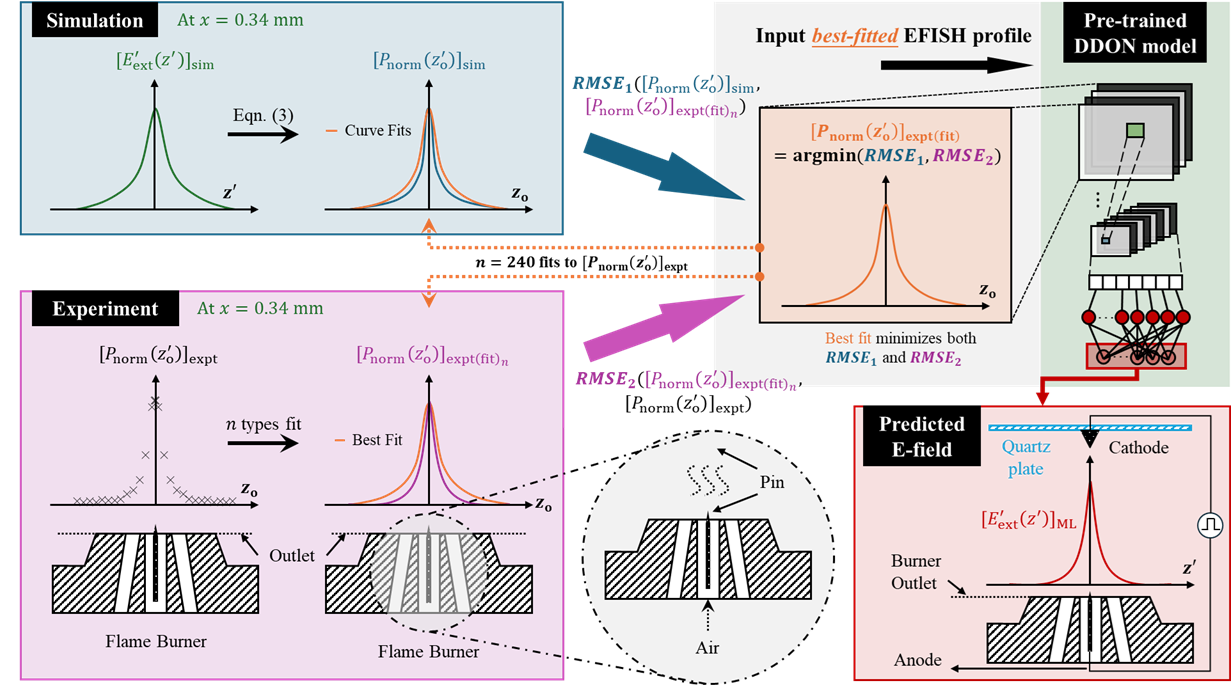}
    \caption{Diagram illustrating the details and flow of the calibration procedure, taking into account the effect of electric the field profile.}
    \label{fig:calibration}
\end{figure}

To facilitate this comparison, we begin by applying a data-smoothing and curve-fitting routine to the experimental E-FISH signal profile (measured under electrostatic conditions), $P^{(2\omega)}_{\text{norm}}(z'_o)_{\text{expt(fit)}}$. This significantly improves the overall quality and spatial resolution of the experimental data by reducing noise and eliminating outliers. In seeking the ‘best’ curve-fit, a total of 8 smoothening methods are first applied to the raw experimental datapoints, $P^{(2\omega)}_{\text{norm}}(z'_o)_{\text{expt}}$ using the \texttt{Smoothdata} function in MATLAB. For each method, 30 window sizes are tested, resulting in a total of 240 unique possible curves. For clarity, we define this group of curves as $[P^{(2\omega)}_{\text{norm}}(z'_o)_{\text{expt(fit)}}]_n$, where $n$ is an integer from 1 to 240. At the same time, equation~(3) is applied to the simulated profile, $E'_{\text{ext}}(z')_{\text{sim}}$, so as to obtain the simulated E-FISH signal profile, $P^{(2\omega)}_{\text{norm}}(z'_o)_{\text{sim}}$. The best curve-fit (from the group of 240 curves) is then defined as the corresponding profile that produces the minimum combined root mean square error (RMSE) relative to both $P^{(2\omega)}_{\text{norm}}(z'_o)_{\text{sim}}$ and $P^{(2\omega)}_{\text{norm}}(z'_o)_{\text{expt}}$ (see figure~\ref{fig:calibration}). For brevity, we simply term this chosen best curve-fit as $P^{(2\omega)}_{\text{norm}}(z'_o)_{\text{expt(fit)}}$, while emphasizing that it is this profile which is fed into the ML model to generate $E'_{\text{ext}}(z')_{\text{ML}}$. Such a detailed scheme enables a certain degree of rigour, and more importantly ensures a systematic and repeatable data-processing procedure. The same curve-fitting routine (yielding the smallest combined RMSE discussed in section~\ref{ml}) is applied to smooth the E-FISH data profiles acquired in the plasma.

To illustrate the differences between these predictions (of $E'_{\text{ext}}(z')_{\text{sim}}$ and $E'_{\text{ext}}(z')_{\text{ML}}$), we generate two calibration curves corresponding to both profiles (captured by $\Gamma_{\text{sim}}$ and $\Gamma_{\text{ML}}$)  as displayed in figure~\ref{fig:cal}. The slope of the resulting lines exhibits a difference of about a factor of 20 \%, and it is this uncertainty which we carry forward when computing the electric fields strengths in the plasma. While good agreement between the electric field profiles predicted by our electrostatic simulations versus DDON (i.e., $E'_{\text{ext}}(z')_{\text{sim}}$ and $E'_{\text{ext}}(z')_{\text{ML}}$) is observed, we have also examined a few possible sources of error that may have arisen in this work. These are discussed in the following section~\ref{subsec:Eerror}.

\subsection{Possible sources of error between $E'_{\text{ext}}(z')_{\text{sim}}$ and $E'_{\text{ext}}(z')_{\text{ML}}$}\label{subsec:Eerror}

We suggest, in order of probability, a few possible reasons discrepancies between the two calibration curves presented in figure~\ref{fig:cal}:
\begin{enumerate}
    \item \textbf{Uncertainty in probe location.} Due to the 10~$\mu$m resolution of the micrometer screw gauge relative to the electrode diameter, the exact interrogation region cannot be precisely determined. This is particularly true along the vertical $x$-axis (with respect to the HV electrode), as well as the transverse ($y$-axis) direction.

    \item \textbf{The choice of curve-fit to the experimental data, $P^{(2\omega)}_{\text{norm}}(z'_o)_{\text{expt(fit)}}$, could potentially introduce some error.} A variation of RMSE within 0.05 or less can result in uncertainties of up to 30\%. This further reinforces the need for a transparent and rigorous curve-fitting procedure.

    \item \textbf{The ML algorithm is not perfect, and could introduce some uncertainty.} Nonetheless, we believe this to be of least influence since previous work~\cite{yang2025deep, Yang2025Interpretable} has shown that excellent reconstruction accuracy can be obtained if E-FISH signal profiles with the appropriate spatial resolution and quality are provided.
\end{enumerate}

\section{Results}
\label{Res}

\subsection{Transient stages of Discharge evolution}
\label{Trans}

\begin{figure}[htbp]
    \centering

        \centering
        \includegraphics[width=\textwidth]{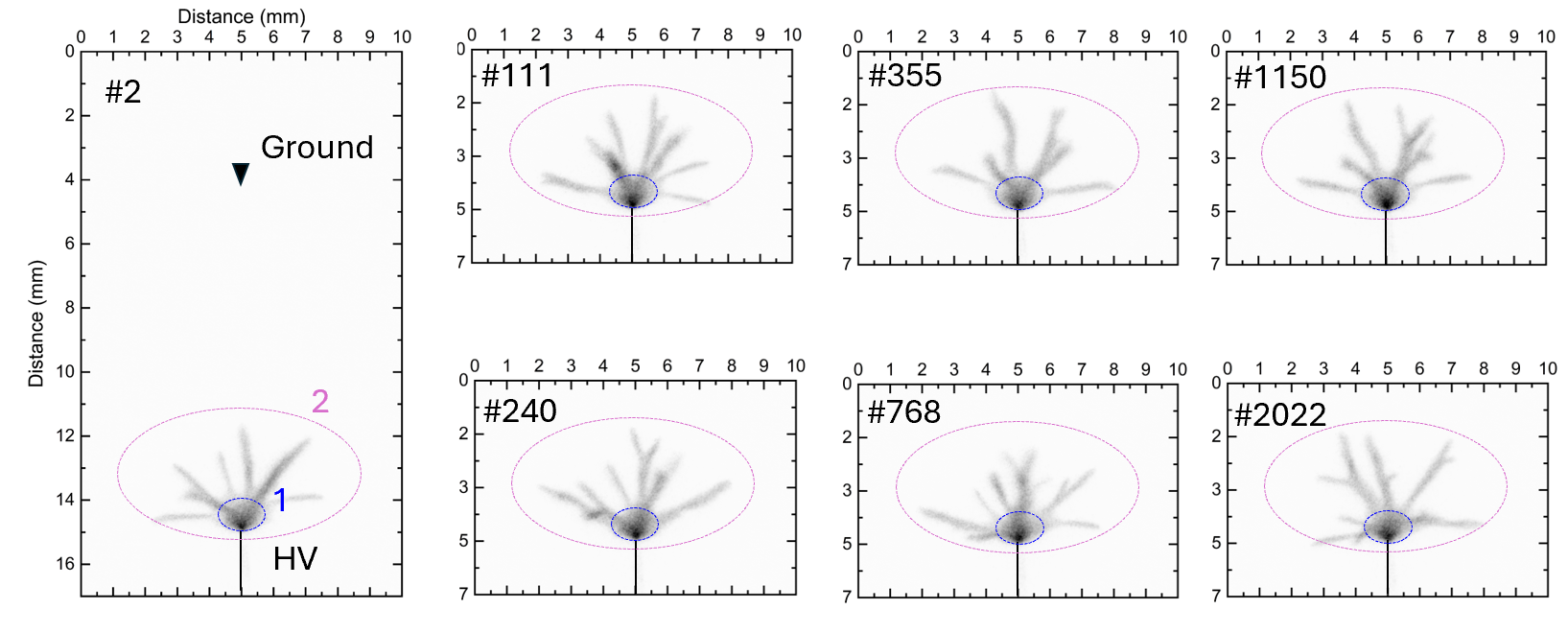}
    \caption{Representative discharge morphology obtained using single-shot iCCD imaging with an integration time of 20 ns. The steady region is indicated in blue, the unsteady region is indicated in purple.}
    \label{fig:discharge}
\end{figure}

To assess the spatial and temporal behavior of the discharge, we begin by examining the series of time-resolved and time-integrated emission data. The first set of results, shown in Figure~\ref{fig:discharge}, presents a selection of 7 images randomly chosen from the full dataset of 2048 single-shot acquisitions. These images correspond to broadband UV emission with a 20~ns integration time, capturing the entire discharge event in each frame.

The discharge morphology is composed of two distinct regions. The first is a central, stable region near the high-voltage (HV) electrode, referred to as region~1 in Fig.~\ref{fig:discharge}, which exhibits a highly reproducible emission pattern across all recorded shots. This steady behavior indicates that the discharge is repeatable and spatially stable within a radius of approximately 1~mm from the HV electrode tip. The second region, referred to as region~2, consists of multiple filamentary streamers that propagate radially outward. Unlike the bulk region, the number, direction, and branching patterns of the streamers vary from shot to shot, with no consistent structure observed over the dataset. This indicates a stochastic streamer formation process in the outer region, extending roughly up to 3~mm from the HV electrode.

\begin{figure}[htbp]
    \centering

        \centering
        \includegraphics[width=\textwidth]{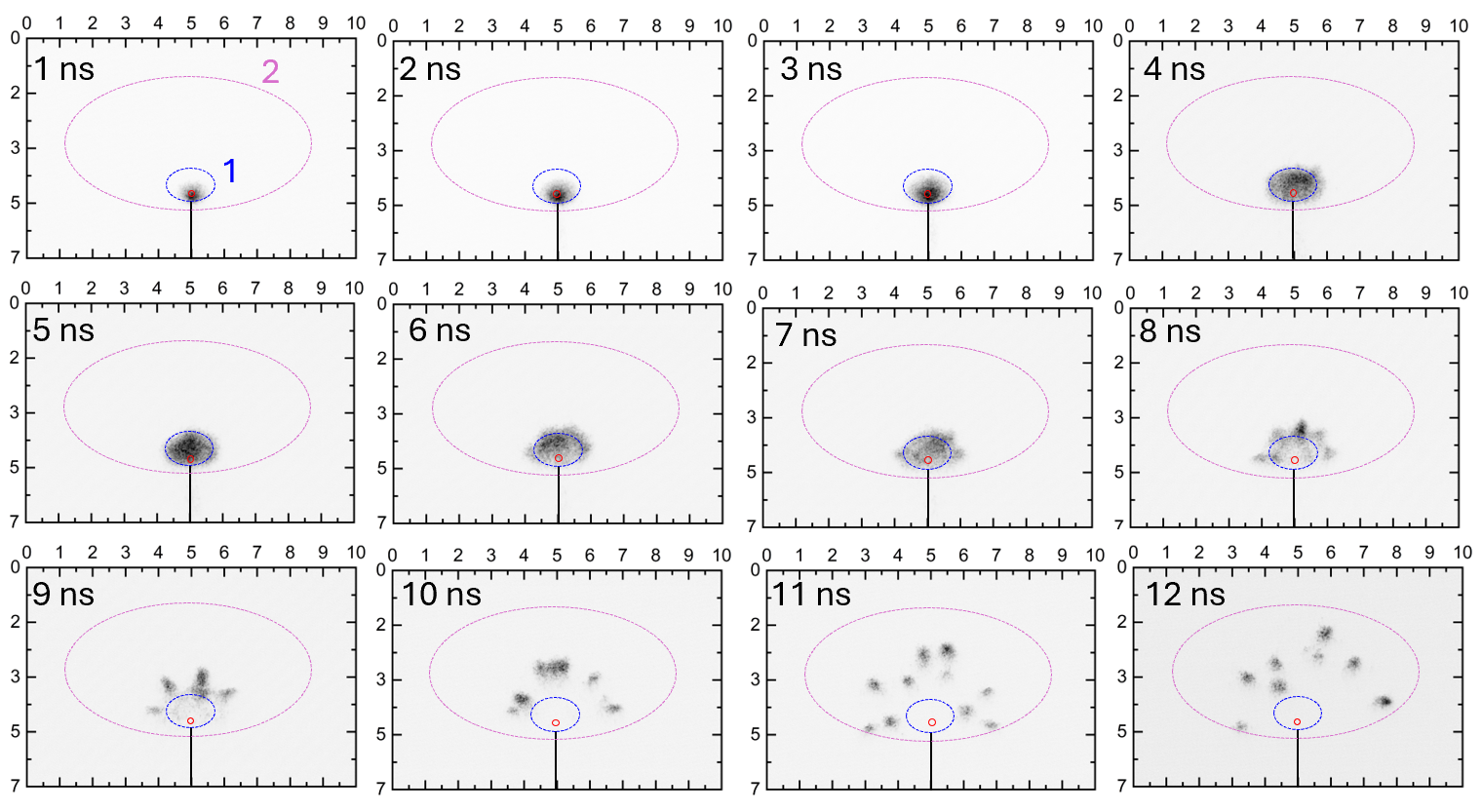}
    \caption{Discharge morphology obtained using single-shot iCCD imaging with an integration time of 1 ns. The red dot indicates the position of the laser. The steady region is indicated in blue, the unsteady region is indicated in purple.}
    \label{fig:timedischarge}
\end{figure}

Figure~\ref{fig:timedischarge} provides further insight into the temporal evolution of the discharge. It shows 12 single-shot images captured similarly with broadband detection, but integrated over a 1~ns gate width. These images are uncorrelated in time (i.e., each is from an independent discharge event), but collectively illustrate the dynamics of the plasma. The same two regions are represented in each frame with a red dot indicating the the laser focus for z=0~mm.

In the early time window (defined as 1–6~ns), the emission remains confined to region~1 and exhibits a highly consistent structure from shot to shot. This phase corresponds to the so-called inception cloud propagation, as commonly described in the literature~\cite{nijdam2020physics}. In the following time interval (6–7~ns), signs of destabilization appear, marking a  transition in the discharge dynamics. From 8~ns onward, streamer filaments begin to emerge and propagate, characterized by increasing spatial extension and stochasticity in region~2. This transition reflects a shift in the discharge regime: while the initial nanoseconds correspond to a quasi-steady corona phase, the subsequent evolution is dominated by filamentary streamer development with inherent variability.

\subsection{Spatiotemporal E-FISH signal}

\begin{figure}[htbp]
    \centering

        \centering
        \includegraphics[width=\textwidth]{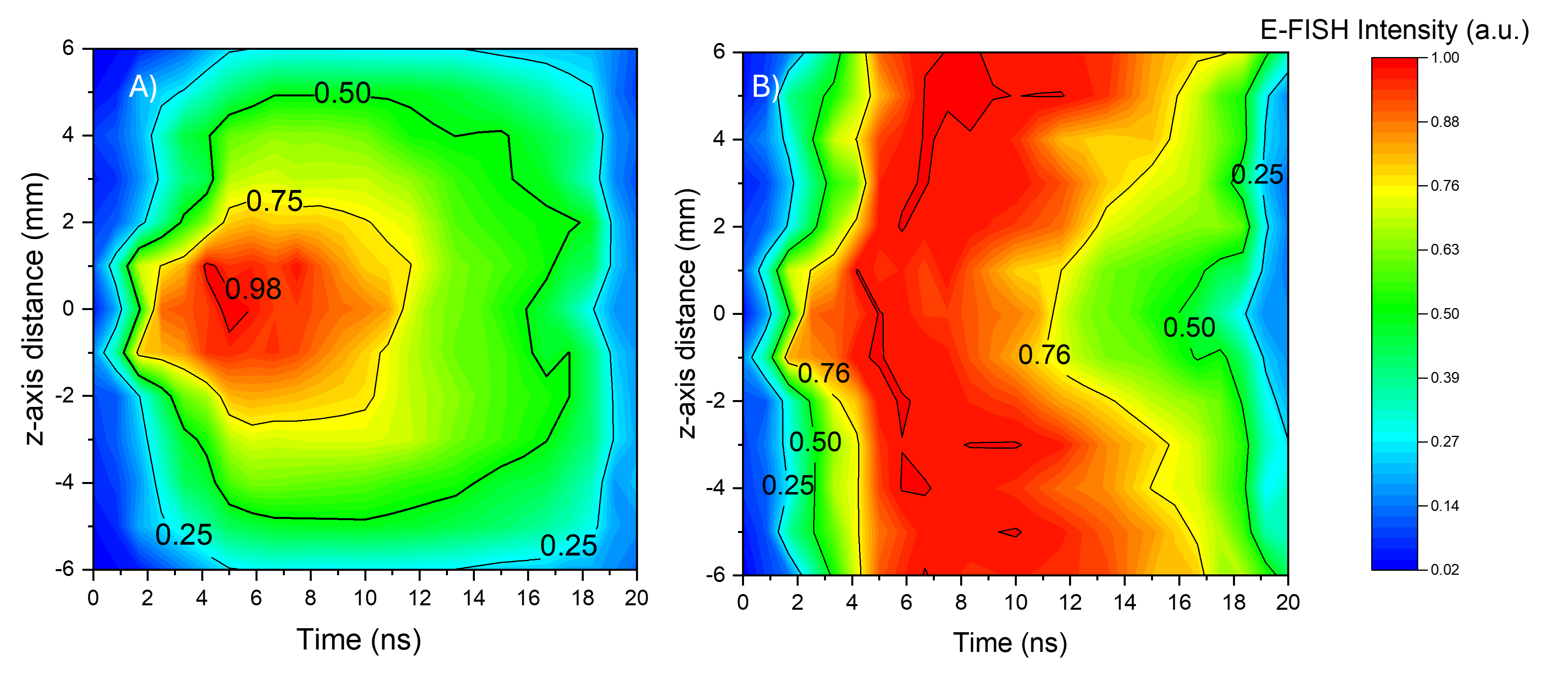}
    \caption{Spatio-temporal maps of the normalized E-FISH signal along the laser axis. (A) Signal normalized by the global maximum of the dataset. (B) Signal normalized locally at each spatial position to highlight relative temporal variations.}
    \label{fig:EFISHxt}
\end{figure}

Figure~\ref{fig:EFISHxt} presents the spatio-temporal evolution of the E-FISH measurements along the laser axis. In Figure~\ref{fig:EFISHxt}.A, the E-FISH signal has been normalized with respect to the global maximum of the entire dataset, while Figure~\ref{fig:EFISHxt}.B shows the same data normalized at each spatial position by its local maximum. These two complementary representations allow for a detailed analysis of both the absolute E-FISH distribution and the local temporal E-FISH dynamics along the discharge axis.

In Figure~\ref{fig:EFISHxt}.A, the highest E-FISH signals are observed close to the high-voltage electrode, with the field reaching its maximum intensity between approximately 5 and 6~ns. This strong, early field enhancement results from the rapid voltage rise and the associated Laplacian amplification at the electrode tip, initiating the formation of the inception cloud~\cite{nijdam2020physics}. During this initial stage, space charge begins to accumulate near the electrode, progressively modifying the local field structure while remaining relatively confined in space. This inception cloud is what we refer to as region~1 in Figure~\ref{fig:timedischarge}. Outside of this region, the E-FISH signal rapidly drops below 75\% of its maximum value. At the limits of our ($\pm$6~mm) measurements, the E-FISH signal has dropped significantly, down to 25\% of the maximum signal.

The normalized map in Figure~\ref{fig:EFISHxt}.B provides further insight into the local temporal evolution of the signal along the axis. In the region near the HV electrode ($\pm$1~mm), the E-FISH signal rises almost simultaneously across positions, confirming that field establishment during the inception phase is spatially synchronized along the axis, consistent with a quasi-stationary inception cloud~\cite{nijdam2020physics}. However, beyond this central region, a slight temporal shift in the E-FISH signal peak becomes apparent (especially when considering the maximum) as the measurement position moves further from the electrode. This delay reflects a combination of physical effects: first, the progressive axial expansion of the inception cloud as ionization spreads outward; and second, at later times ($>$ 5--7~ns), the onset of streamer activity intersecting the laser axis, as confirmed by the time-resolved data in Figure~\ref{fig:timedischarge}. These streamer heads, while stochastically located from shot to shot, begin to appear in the laser measurement domain during the second half of the voltage pulse, leading to additional local signal enhancement contributions.

Overall, the E-FISH measurements capture the highly reproducible field buildup associated with the inception phase near the high-voltage electrode, as well as the progressive axial expansion of the field during the early development of the discharge. At larger distances and later times, the observed temporal shifts suggest that local field enhancement may also reflect the initial stages of streamer head formation but becomes challenging to interpret.

\subsection{Time-resolved electric field reconstruction}

\begin{figure}[htbp]
    \centering
    \includegraphics[width=\textwidth]{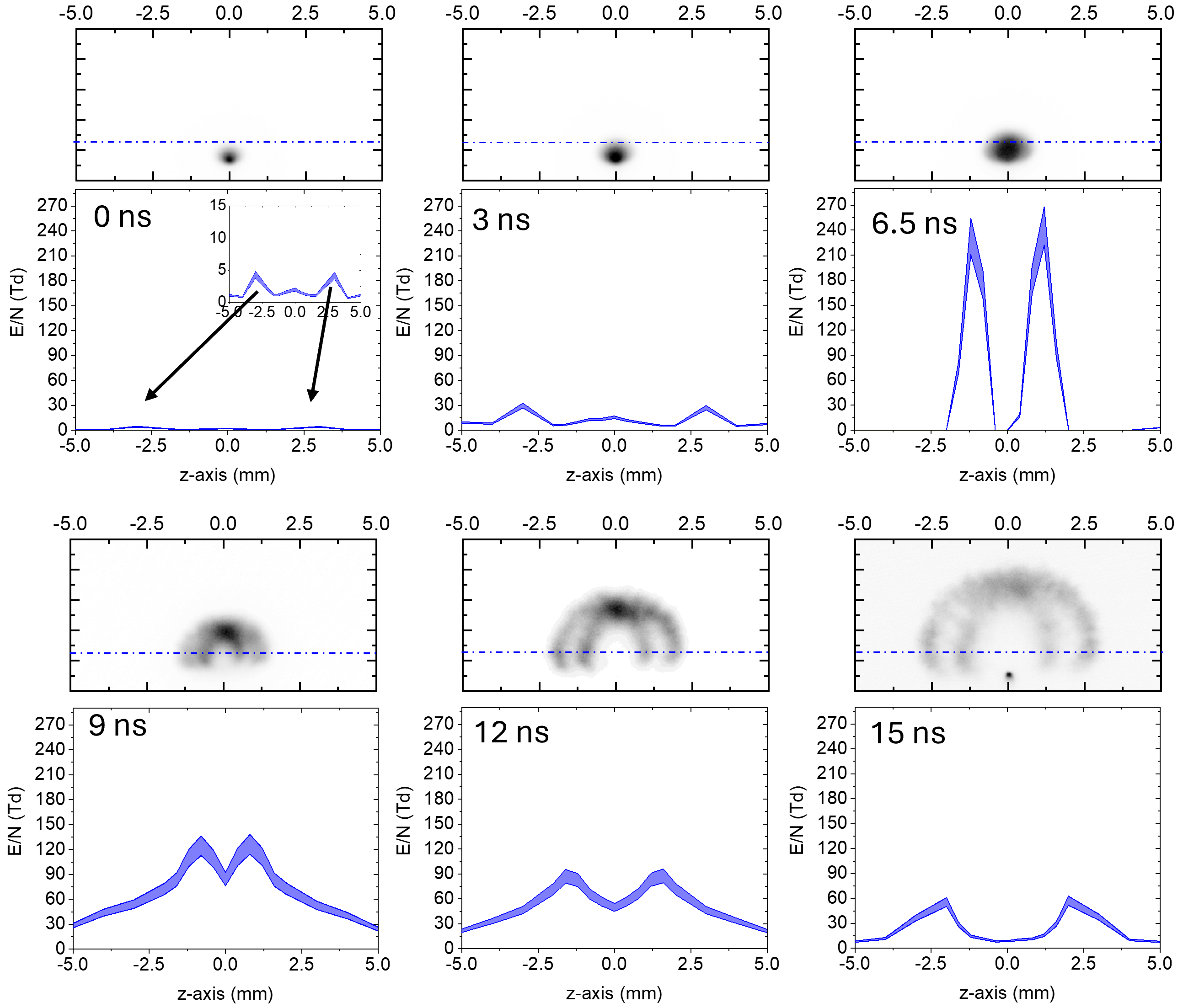}
    \caption{Representative time-resolved electric field profiles reconstructed
    using the DDON-based inversion model. Each panel couples the E-field profile
    (in units of Townsend) with its matching iCCD emission image at the same time delay. The dashed
    horizontal line indicates the axial location probed by E-FISH. Blue curves correspond to the
    two calibration methods described in Section~\ref{calibration}, with the
    shaded area denoting their envelope.}
    \label{fig:efield_reconstruction}
\end{figure}

Figure~\ref{fig:efield_reconstruction} presents representative time steps of the
electric field reconstructed using the DDON-based inverse model. For each delay,
the E-field profile along the laser axis (expressed in units of Townsend) is shown together with the corresponding iCCD image of the
discharge, both accumulated over 2048 shots. The dashed horizontal line in the
iCCD images marks the position of the E-FISH measurement. The two blue curves in
each subplot correspond to the calibration procedures discussed in
Section~\ref{calibration}, and the shaded region between them represents the
uncertainty bounds within which the actual electric field lies.

Three distinct behaviors can be identified in the temporal evolution of the
electric field.

\paragraph{(1) Early inception cloud growth (0--6~ns).}
During the initial nanoseconds, the iCCD images show the development of a
compact inception cloud near the high-voltage electrode. The probed position,
however, lies well above this region, making it difficult to infer the local
field structure. The reconstructed field profiles exhibit
two symmetric peaks located more than 2.5~mm from the electrode tip. Although
the spatial interpretation of these peaks is limited by the large separation
between the probe volume and the emission region, the magnitude of the electric
field evolves rapidly during this stage, increasing from approximately
5~Td to more than 240~Td by 6~ns. This trend is consistent with the fast rise of
the applied voltage and the onset of ionization within the inception cloud,
which enhances the local susceptibility of the medium and contributes to the
observed growth in the E-FISH signal.

\paragraph{(2) Shell-like structure formation (around 6.5~ns).}
At 6.5~ns, the discharge morphology undergoes a qualitative transformation:
iCCD images reveal the emergence of a shell-like structure surrounding the
inception region. The reconstructed electric field shows its highest amplitudes
at this time, with two pronounced peaks located at the boundaries of the
shell-like emission region. Such structures are consistent with the accumulation
of space charge at the expanding plasma boundary, which can strongly reshape the
local electric field even before streamer formation begins, as reported in prior
numerical studies~\cite{naidis2005conditions,luque2012density}. The observed
field magnitudes (up to 250~Td) fall within the typical range required for
processes such as electron-impact ionization and photoionization, supporting the
interpretation that the discharge is approaching the destabilization threshold.
Because the shell position is remarkably repeatable across shots, the field
measurements at this time are more directly representative of a well-defined
plasma boundary compared to earlier or later stages.

\paragraph{(3) Destabilization and streamer formation (9--15~ns).}
At later times, the shell destabilizes and streamer formation begins. The iCCD
images show the progressive emergence of multiple off-axis streamer filaments,
each contributing to a highly non-axisymmetric discharge morphology. The
reconstructed electric field profiles retain two dominant peaks, which gradually
separate in accordance with the expanding radial structure visible in the iCCD
images. However, these peaks can no longer be interpreted as the local field at
a specific streamer head. Instead, they represent the superposition of the
electric fields generated by many streamer filaments at different azimuthal
positions. Single-shot iCCD images confirm substantial shot-to-shot variability,
and since the DDON model typically assumes axisymmetry, the reconstruction at these
times reflects only an averaged projection of the true three-dimensional field
distribution. Furthermore, as the discharge evolves, the streamer heads do not
necessarily occupy the same spatial location in every shot, reducing the
physical representativeness of the averaged field at the probed position.

\vspace{0.2cm}

Overall, the DDON-based E-FISH reconstruction reliably captures the strong
electric-field enhancement associated with the formation of the shell-like
structure and its subsequent destabilization.

\section{Discussion}
\label{Disc}

\subsection{Applicability of axisymmetric inversion in the presence of streamer stochasticity}
\label{stocha}

\begin{figure}[htbp]
    \centering

        \centering
        \includegraphics[width=\textwidth]{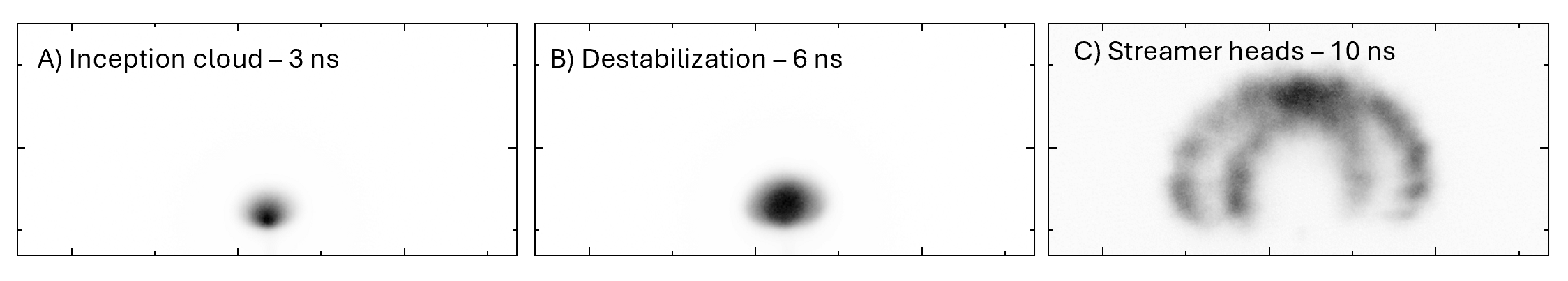}
    \caption{Discharge morphology at selected time instants obtained using  broadband iCCD emission imaging averaged over 2048 shots with an integration time of 1 ns.}
    \label{fig:streamer_symmetry}
\end{figure}

For accurate reconstruction, the DDON-based electric field inversion algorithm used in this work generally requires symmetric electric field distributions along the direction of laser propagation about the vertical x-axis. This is primarily due to the symmetric features of the profiles within the model's training dataset. We justify this condition on the basis that during the early inception phase, ionization remains localized and axisymmetrically distributed near the electrode tip (see Figure~\ref{fig:streamer_symmetry}A). During this stage, as mentioned earlier in section~\ref{Trans} , the discharge morphology remains highly stable and reproducible between pulses, enabling accurate inversion of the E-FISH signals into local electric field profiles.

As the discharge evolves toward streamer formation, axisymmetry is progressively lost on a \textbf{single-shot} basis due to stochastic streamer branching and propagation. This transition is clearly seen in time-resolved iCCD images (Figure~\ref{fig:timedischarge}). In Figure~\ref{fig:streamer_symmetry}A and \ref{fig:streamer_symmetry}B, the discharge remains confined to region 1 and highly symmetric both \textbf{on average} and from shot-to-shot, while in Figure~\ref{fig:streamer_symmetry}C, the averaged emission shows a new structure, symmetric "on average" but likely non-symmetric from shot-to-shot, formed in region~2 by the streamer heads.

This extensive averaging over 2048 shots partially restores statistical symmetry in the E-FISH signals (Figure~\ref{fig:EFISHxt}B), smoothing out the stochastic streamer locations from pulse to pulse. This results in stable signal contours that approximate an axisymmetric field evolution. While this statistical symmetry results in accurate profile inversion using the DDON model even beyond the inception phase, these reconstructed profiles must be interpreted with caution.

First, regarding the inversion itself, an important limitation arises from the nonlinear nature of the E-FISH signal once field axisymmetry is no longer perfectly valid. Because the E-FISH response scales with the square of the local electric field, the contribution of each single-shot realization to the averaged signal depends sensitively on whether, and where, a streamer head intersects the laser axis in a given shot. Consequently, the averaged E-FISH signal at each position represents a \textit{nonlinearly} weighted sum over many discharge realizations, where on-axis streamer events can disproportionately influence the result.
Second, with respect to the initial objective of this work (to obtain quantitative electric field measurements enabling the interpretation of inception cloud dynamics and destabilization) the relevant information is primarily contained in the 0–6~ns time window, where the discharge remains symmetric and unambiguous. For this reason, our analysis focuses on time instants prior to streamer branching.

Nevertheless, the subsequent results (7~ns onward) remain meaningful when interpreted appropriately. Although they no longer describe the local field of the cloud boundaries, they still broadly reflect the collective electric environment near the high-voltage electrode and, when analyzed together with iCCD imaging, can provide valuable insight into streamer propagation and global discharge evolution.




\subsection{Validation of the reconstructed electric field}

\begin{figure}[htbp]
    \centering
    \includegraphics[width=\textwidth]{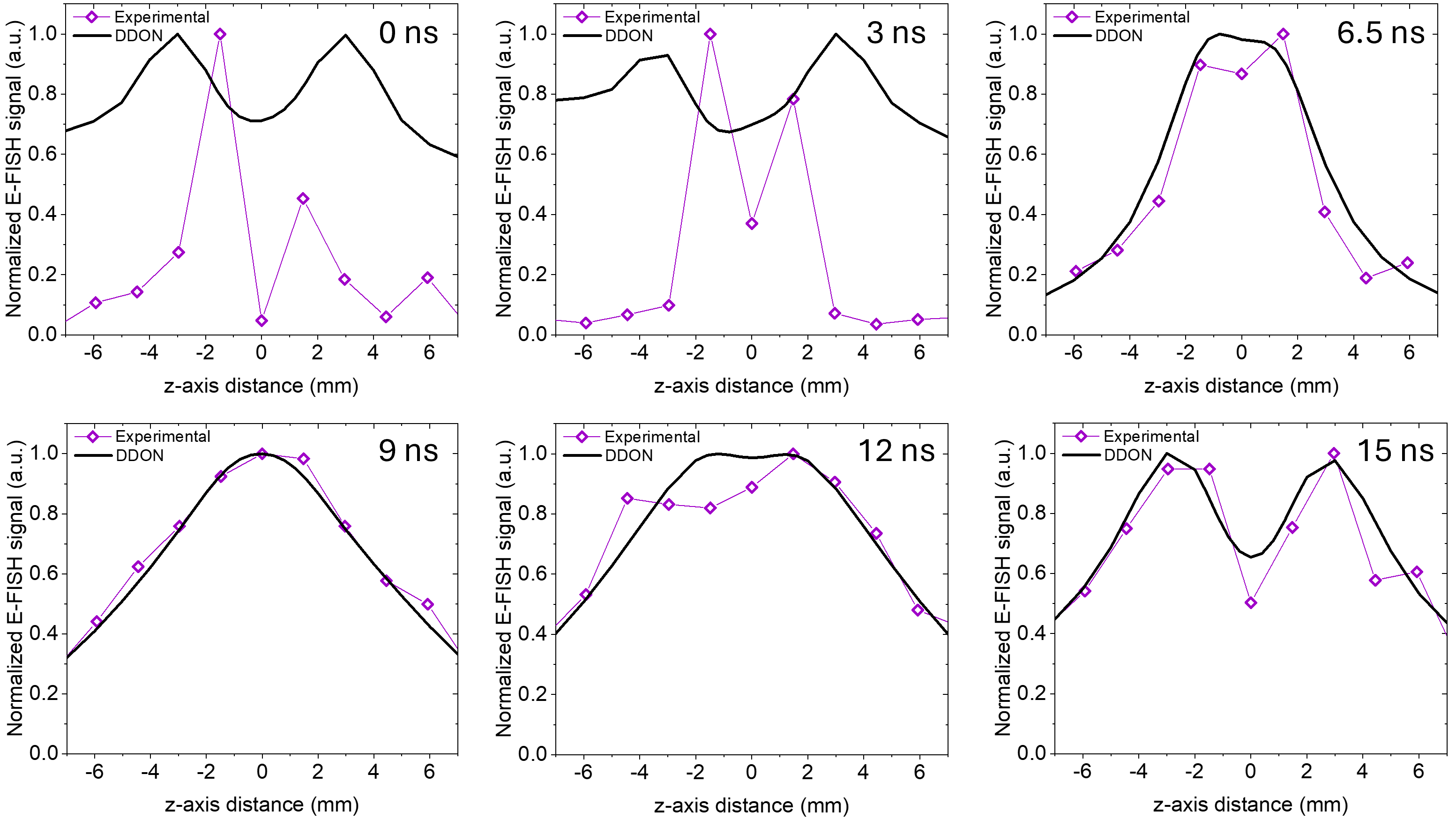}
    \caption{Comparison between experimentally measured E-FISH signals (purple)
    and forward-calculated E-FISH profiles obtained by applying Eq.~(3) to the
    reconstructed electric fields. Good agreement indicates successful
    reproduction of the measured signal by the DDON-based inversion.}
    \label{fig:efish_validation}
\end{figure}

To assess the reliability of the DDON-based reconstruction, the experimentally
measured E-FISH signals were compared with forward-calculated signals obtained
by solving Eq.~(3) using the reconstructed electric fields as input. Since a
given E-FISH intensity profile corresponds to a unique axial electric-field
distribution under the assumptions of Section~\ref{calibration}, good agreement
between the measured and forward-calculated profiles constitutes a strong
consistency check for the inversion procedure. Representative comparisons are
shown in Figure~\ref{fig:efish_validation}, where the experimental profiles
(purple curves) are plotted against their forward-computed counterparts.

A clear trend emerges across the investigated time delays. From approximately
6~ns onward, the agreement between the measured and forward-calculated E-FISH
profiles is excellent. This includes the period associated with the formation of
the shell-like structure (around 6.5~ns) as well as the subsequent destabilized
phase in which streamer filaments begin to propagate. Even in this latter
regime, where non-axisymmetric features develop on a single-shot basis, the forward model reproduces the experimental E-FISH profiles with
high fidelity. This strongly supports the validity of the reconstructed electric
fields for delays between 6 and 15~ns, indicating that the DDON inversion
successfully captures the field magnitude and spatial variation relevant to both
the shell boundaries and their early destabilization.

In contrast, at earlier times (0--5~ns), noticeable discrepancies arise between
the experimental signals and the forward-calculated profiles. These mismatches
limit confidence in the reconstructed field shapes presented for the 0 and 3~ns
time steps. Several factors may contribute to this reduced agreement.

First, small uncertainties in electrode alignment,
timing jitter ($<$ 3 ns as described in section \ref{EFD}), or focal positioning can strongly affect the E-FISH signal when
the emitting region is compact and evolving rapidly. Second, the very steep
intensity gradients typically observed during the initial nanoseconds of the
discharge, where the normalized E-FISH signal transitions from near zero to its
maximum over a short axial distance, can amplify reconstruction errors, since the
inversion is highly sensitive to the precise spatial location of these sharp
features. Third, during the early phase the probing laser intersects the region
well above the inception cloud. As a result, the measured E-FISH signal reflects
field variations far from the plasma boundary rather than the field directly
adjacent to the ionization front. This geometric mismatch reduces the amount of
field-relevant information available to the inversion algorithm in the earliest
time steps.

Addressing these limitations will require acquiring E-FISH measurements along
the discharge axis (i.e.\ translating the probe beam in the $x$-direction)
during the inception stage, and with minimal timing jitter. Such measurements would provide direct access to the
field distribution at the evolving plasma boundary, enabling a more complete
characterization of inception-cloud propagation and improving the robustness of
the inversion during the first few nanoseconds.

Overall, the validation confirms that the reconstructed electric fields are
reliable from the formation of the shell-like structure onward, while the early
inception-phase reconstructions should be interpreted with caution due to
geometric, physical, and methodological constraints intrinsic to the present
measurement configuration.

As a final note, we briefly examine how the inclusion of ML-assisted reconstruction affects the predicted electric field strengths. As shown in  Figure~\ref{fig:cal}C), the peak field strengths obtained via our DDON model are about 30\% higher than estimates obtained with regular (single-point) calibration. Likewise, the drop in field strength at 7.5 ns (likely due to the double-peak profiles) also leads to an overestimate by the uncorrected values. These observations are in good agreement with \cite{chng2022effect}, who observed that accounting for the entire electric field distribution provided more accurate estimates of the field strength, but in general, did not drastically alter the overall shape of the time evolution curve at a particular point in space.

\subsection{Physical interpretation of the inception-to-streamer transition}

The E-FISH measurements and DDON reconstructions presented above indicate that the
most physically reliable time window corresponds to the formation and destabilization
of the shell-like structure between 6 and 7~ns. Which correspond to our initial intent of characterizing the inception cloud destabilization. Both the validation results and the
iCCD imaging confirm that this interval combines (i) sufficient symmetry for a robust
inversion and (ii) clear physical signatures associated with positive streamer physics.
Therefore, the discussion in this section focuses primarily on this regime.

Around 6--7~ns, the reconstructed electric field reaches its maximum value, with peak
reduced fields of approximately $250~\mathrm{Td}$. This moment coincides with the
formation of a shell surrounding the inception cloud. The spatial structure
and timing of this shell agree well with theoretical descriptions of the transition from
a diffuse inception cloud to individual streamer heads. In particular, Naidis
\cite{naidis2005conditions} shows that once space-charge
accumulation near the ionization front enhances the local electric field sufficiently,
the inception cloud becomes unstable and breaks into discrete ionization fronts driven
by photoionization. Similarly, Luque and Ebert \cite{luque2012density}
demonstrate in their density-model simulations that a radially expanding ionization
front becomes susceptible to perturbations when the space-charge layer develops a
pronounced curvature and the electric field peaks at the front boundary. Their 3D
analysis of streamer branching further shows that this transition is accompanied by
a rapid increase in off-axis Fourier modes, indicating symmetry breaking of the
initially quasi-spherical structure.

Our measurements support this interpretation. At 6.5~ns, the E-field profile displays
two strong peaks symmetric about the axis, matching the boundaries of the shell observed in the iCCD images. The magnitude of the reduced field ($\sim 250$~Td)
falls squarely within the regime where electron-impact ionization, excitation of
nitrogen states, and photoionization all become highly efficient. This is consistent
with the physical mechanisms sustaining positive streamer propagation. In particular:

\begin{itemize}
    \item Electron-impact ionization increases rapidly above $\sim 150$--$200$~Td, so a
    peak value of $250$~Td ensures substantial ionization growth ahead of the front.
    \item In this field range, electron-impact excitation of nitrogen states that emit photoionizing UV radiation (98–102.5 nm) becomes efficient, providing the source terms required for nonlocal photoionization. This mechanism, central to positive streamer propagation, is incorporated in the photoionization models used by Naidis~\cite{naidis2005conditions} and others.
    \item The combined effect leads to nonlocal electron production capable of sustaining
    ionization fronts even in regions where the local field would otherwise be insufficient.
\end{itemize}

\begin{figure}[htbp]
    \centering
    \includegraphics[width=0.75\linewidth]{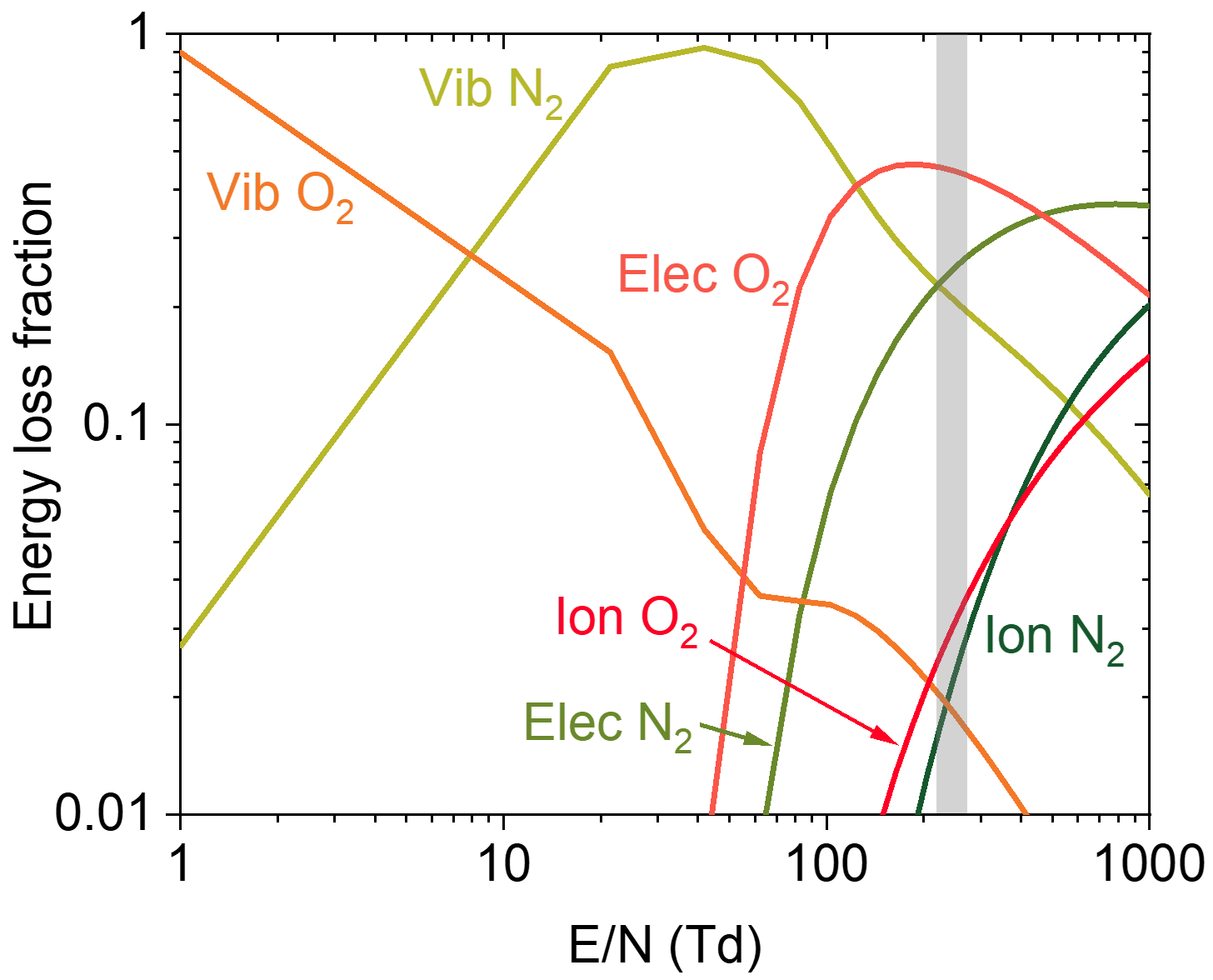}
    \caption{Electron energy loss fractions in atmospheric-pressure air as a function
    of the reduced electric field $E/N$ (calculated with the set of cross sections from~\cite{phelps1985anisotropic, lawton1978excitation}). The shaded grey region indicates the range
    of field strengths reconstructed in this work ($\sim 230$--$270~\mathrm{Td}$).}
    \label{fig:energyloss}
\end{figure}

Figure~\ref{fig:energyloss} places these field strengths in the context of electron–impact collision processes calulated using the set of cross sections from~\cite{phelps1985anisotropic, lawton1978excitation}. The shaded band highlights the
range $E/N \approx 230$--$270~\mathrm{Td}$, corresponding to the reconstructed
fields during formation and destabilization of the shell-like structure. In this
interval, electron-impact electronic excitation of O$_2$ represents the
dominant inelastic channel (about 45\% of the total electron energy loss), while
electronic excitation of N$_2$ and vibrational excitation of N$_2$
contribute approximately 25\% and 20\%, respectively. Only a minor fraction of
the electron energy is dissipated through ionization. This partitioning indicates
that the discharge operates in a regime where excitation processes, and in
particular those participating in photoionizing transitions in N$_2$ and subsequent
ionization of O$_2$, are highly efficient, whereas impact ionization is already
active but not yet predominant. The measured fields are therefore consistent with
the hypothesis in which a strongly excited, space-charge-dominated
inception cloud reaches the conditions for destabilization and streamer
formation, as described in~\cite{naidis2005conditions,luque2012density, nijdam2020physics}.

After 7~ns, the measured E-field amplitude begins to decrease, and the iCCD images
show clear formation and propagation of streamer heads. This behavior is consistent
with the onset of streamer instabilities described by Luque and Ebert, where the
shell structure loses its symmetry and breaks into discrete branches once the ionization
front becomes sufficiently thin and space-charge-induced field enhancement concentrates
at localized points. Although the DDON reconstruction becomes less certain in this
highly non-axisymmetric regime, the observed decrease of field at the laser position is
expected: the highest fields are now located at the tips of individual streamer heads
propagating away from the probe volume.

Taken together, the measured peak fields, the temporal evolution of the shell-like
structure, and the subsequent branching behavior are consistent with theoretical and
numerical models of positive streamer inception. The field values around 6--7~ns
represent the transition point where the inception cloud develops enough space charge,
enhanced local field, and excitation-driven photoionization to destabilize and break
into propagating streamer channels.

\section{Conclusion}
\label{Conc}


This work demonstrates a methodology and detailed procedure for quantitatively resolving the electric field strength in a nanosecond positive corona discharge. The approach combines both time-resolved E-FISH measurements together with an ML-based (DDON) field reconstruction model, supported by iCCD imaging. The DDON model affords accurate recovery of the underlying electric field distribution from a given E-FISH profile, through a spatial scan of the discharge relative to the laser beam along its optical axis. However, since the corona discharge has an intrinsic tendency to develop stochastically, this scanning procedure risks predicting erroneous electric field information from E-FISH signals which have essentially been averaged over many laser shots and different discharge events. Incorporating both single-shot and time-averaged iCCD imaging therefore serves as a complementary, but crucial, tool for interpreting the validity of these reconstructed electric field profiles.

We successfully apply this method for characterizing the inception cloud and its subsequent destabilization near the high voltage electrode during the nascent stages of the discharge. Accurate quantitative reconstruction of the field requires conditions in which the discharge remains sufficiently axisymmetric and repeatable. These conditions are partially fulfilled during the inception phase, where, starting from 6 ns, the emission pattern is stable from shot to shot and the E-FISH signal closely satisfies the assumptions underlying the DDON model. In this time interval, the reconstructed fields show a rapid increase synchronized with the voltage rise and early ionization dynamics, culminating in peak reduced fields just prior to shell destabilization. Quantitatively, the measured electric-field norm reaches peak reduced values of E/N $\simeq$ 250 Td, with reconstructed fields consistently lying within $\simeq$ 230–270 Td across the shell boundaries, defining the characteristic electric-field scale associated with inception cloud destabilization. These measurements provide physically interpretable, high-fidelity insight into the structure and evolution of the inception cloud.

Beyond approximately 7 ns, the emergence of stochastic streamer filaments introduces increasing asymmetry in the discharge. Although averaging over thousands of realizations restores partial statistical symmetry, the nonlinear nature of the E-FISH response imposes intrinsic limitations on quantitative inversion in this regime. Consequently, reconstructed fields at later times should be interpreted as spatially averaged representations of the local electric environment rather than direct measurements of individual streamer heads. In this regime, the reported field magnitudes represent an effective, ensemble-averaged electric-field norm rather than the instantaneous peak field at a single streamer tip.

Overall, this study establishes the conditions under which DDON-assisted E-FISH inversion yields reliable quantitative field information, with iCCD imaging proving vital for identifying the onset of streamer branching as the principal limit of applicability. Extending E-FISH measurements along the inter-electrode gap will enable full mapping of the inception region, while future diagnostic developments capable of single-shot electric-field resolution will be essential for capturing the inherently stochastic dynamics of streamer propagation.

%
%

\ack{This study was funded by the King Abdullah University of Science and Technology (KAUST), under the grant number BAS/1/1396-01-01.}



\clearpage

\bibliographystyle{iopart-num}
\bibliography{Bib-base}

\providecommand{\newblock}{}
\begin{thebibliography}{10}
\expandafter\ifx\csname url\endcsname\relax
  \def\url#1{{\tt #1}}\fi
\expandafter\ifx\csname urlprefix\endcsname\relax\def\urlprefix{URL }\fi
\providecommand{\eprint}[2][]{\url{#2}}

\bibitem{wang2020nanosecond}
Wang D and Namihira T 2020 {\em Plasma Sources Science and Technology\/} {\bf 29} 023001

\bibitem{lacoste2023flames}
Lacoste D~A 2023 {\em Proceedings of the Combustion Institute\/} {\bf 39} 5405--5428

\bibitem{starikovskaia2014plasma}
Starikovskaia S 2014 {\em Journal of Physics D: Applied Physics\/} {\bf 47} 353001

\bibitem{popov2016kinetics}
Popov N 2016 {\em Plasma Sources Science and Technology\/} {\bf 25} 043002

\bibitem{kolb2008streamers}
Kolb J, Joshi R, Xiao S and Schoenbach K 2008 {\em Journal of Physics D: Applied Physics\/} {\bf 41} 234007

\bibitem{bardos2010cold}
B{\'a}rdos L and Bar{\'a}nkov{\'a} H 2010 {\em Thin solid films\/} {\bf 518} 6705--6713

\bibitem{van2008evaluation}
Van~Heesch E, Winands G and Pemen A 2008 {\em Journal of Physics D: Applied Physics\/} {\bf 41} 234015

\bibitem{nijdam2020physics}
Nijdam S, Teunissen J and Ebert U 2020 {\em Plasma Sources Science and Technology\/} {\bf 29} 103001

\bibitem{takiyama1986measurement}
Takiyama K, Usui T, Kamiura Y, Fujita T, Oda T and Kawasaki K 1986 {\em Japanese journal of applied physics\/} {\bf 25} L455

\bibitem{kuraica1997electric}
Kuraica M and Konjevi{\'c} N 1997 {\em Applied physics letters\/} {\bf 70} 1521--1523

\bibitem{obrusnik2018electric}
Obrusn{\'\i}k A, B{\'\i}lek P, Hoder T, {\v{S}}imek M and Bonaventura Z 2018 {\em Plasma Sources Science and Technology\/} {\bf 27} 085013

\bibitem{bilek2018electric}
B{\'\i}lek P, Obrusn{\'\i}k A, Hoder T, {\v{S}}imek M and Bonaventura Z 2018 {\em Plasma Sources Science and Technology\/} {\bf 27} 085012

\bibitem{bonaventura2011electric}
Bonaventura Z, Bourdon A, Celestin S and Pasko V~P 2011 {\em Plasma Sources Science and Technology\/} {\bf 20} 035012

\bibitem{paris2005intensity}
Paris P, Aints M, Valk F, Plank T, Haljaste A, Kozlov K and Wagner H 2005 {\em Journal of Physics D: Applied Physics\/} {\bf 38} 3894

\bibitem{simeni2018electricz}
Simeni M~S, Tang Y, Frederickson K and Adamovich I 2018 {\em Plasma Sources Science and Technology\/} {\bf 27} 104001

\bibitem{chng2019electric}
Chng T~L, Orel I, Starikovskaia S and Adamovich I 2019 {\em Plasma Sources Science and Technology\/} {\bf 28} 045004

\bibitem{chng2019electricz}
Chng T~L, Brisset A, Jeanney P, Starikovskaia S, Adamovich I and Tardiveau P 2019 {\em Plasma sources science and technology\/} {\bf 28} 09LT02

\bibitem{orr2020measurements}
Orr K, Tang Y, Simeni M~S, Van Den~Bekerom D and Adamovich I~V 2020 {\em Plasma Sources Science and Technology\/} {\bf 29} 035019

\bibitem{adamovich2020nanosecond}
Adamovich I~V, Butterworth T, Orriere T, Pai D~Z, Lacoste D~A and Cha M~S 2020 {\em Journal of Physics D: Applied Physics\/} {\bf 53} 145201

\bibitem{simeni2018electric}
Simeni M~S, Tang Y, Hung Y~C, Eckert Z, Frederickson K and Adamovich I~V 2018 {\em Combustion and Flame\/} {\bf 197} 254--264

\bibitem{goldberg2018electric}
Goldberg B~M, Chng T~L, Dogariu A and Miles R~B 2018 {\em Applied Physics Letters\/} {\bf 112}

\bibitem{lepikhin2020electric}
Lepikhin N, Luggenh{\"o}lscher D and Czarnetzki U 2020 {\em Journal of Physics D: Applied Physics\/} {\bf 54} 055201

\bibitem{chng2020electric}
Chng T~L, Starikovskaia S~M and Schanne-Klein M~C 2020 {\em Plasma Sources Science and Technology\/} {\bf 29} 125002

\bibitem{yang2025deep}
Yang Z, Sugeng E~S and Chng T~L 2025 {\em Plasma Sources Science and Technology\/}

\bibitem{Yang2025Interpretable}
Yang Z, Sugeng E~S, Alicherif M and Chng T~L 2025  (\textit{Preprint} \eprint{2512.00359}) \urlprefix\url{https://arxiv.org/abs/2512.00359}

\bibitem{alkhalifa2024quantifying}
Alkhalifa A~M, Di~Sabatino F, Steinmetz S~A, Pfaff S, Huang E, Frank J~H, Kliewer C~J and Lacoste D~A 2024 {\em Journal of Physics D: Applied Physics\/} {\bf 57} 385204

\bibitem{chng2022effect}
Chng T~L, Pai D~Z, Guaitella O, Starikovskaia S~M and Bourdon A 2022 {\em Plasma Sources Science and Technology\/} {\bf 31} 015010

\bibitem{nakamura2021electric}
Nakamura S, Sato M, Fujii T, Kumada A and Oishi Y 2021 {\em Physical Review A\/} {\bf 104} 053511

\bibitem{nakamura2022measurement}
Nakamura S, Sato M, Fujii T and Kumada A 2022 Measurement method for electric field in streamer discharge based on electric-field-induced second-harmonic generation {\em 2022 IEEE International Conference on Plasma Science (ICOPS)\/} (IEEE) pp 1--1

\bibitem{limburgnumerical}
Limburg A, Guo Y, Nijdam S, Ellenbroek W and Toschi F

\bibitem{guo2025measurement}
Guo Y, Limburg A, Laarman J, Teunissen J and Nijdam S 2025 {\em Physical Review Research\/} {\bf 7} 013051

\bibitem{chen2024measurement}
Chen S, He H, Chen Y, Liu Z, Xie S, Che J, He K and Chen W 2024 {\em Measurement\/} {\bf 231} 114576

\bibitem{naidis2005conditions}
Naidis G 2005 {\em Journal of Physics D: Applied Physics\/} {\bf 38} 2211

\bibitem{luque2012density}
Luque A and Ebert U 2012 {\em Journal of Computational Physics\/} {\bf 231} 904--918

\bibitem{phelps1985anisotropic}
Phelps A and Pitchford L 1985 {\em Physical Review A\/} {\bf 31} 2932

\bibitem{lawton1978excitation}
Lawton S and Phelps A 1978 {\em The Journal of Chemical Physics\/} {\bf 69} 1055--1068

\end{thebibliography}

\end{document}